\documentstyle[epsfig,psfig]{texas}

\def\apj{ApJ}
\def\baas{BAAS}
\def\mnras{MNRAS}
\def\aap{A\&A}

\begin{document}

\def\bline{\rule[1.2mm]{3em}{0.1mm}}
\def\msun{M_\odot}
\def\zsun{Z_\odot}
\def\lsun{L_\odot}
\def\fun#1#2{\lower3.6pt\vbox{\baselineskip0pt\lineskip.9pt
  \ialign{$\mathsurround=0pt#1\hfil##\hfil$\crcr#2\crcr\sim\crcr}}}
\def\la{\mathrel{\mathpalette\fun <}}
\def\ga{\mathrel{\mathpalette\fun >}}
\def\eg{{\it e.g., }}
\def\etal{{\it et al. }}
\def\etalc{{\it et al., }}
\def\pc {{\rm pc}}
\def\mpc{{\rm Mpc}}
\def\kpc{{\rm kpc}}
\def\Mpc{{\rm Mpc}}

\def\he#1{\hbox{${}^{#1}{\rm He}$}}
\def\li#1{\hbox{${}^{#1}{\rm Li}$}}

\def\imfm{\xi_{\star}}
\def\mrem{m_{\rm rem}}
\def\avg#1{\langle #1 \rangle}
\def\sigbar{\avg{\sigma}}
\def\rbar{\hbox{$\avg{r}$}}
\def\omegam{\Omega_{\rm Macho}}
\def\omegab{\Omega_{\rm B}}
\def\omegalya{\Omega_{{\rm Ly}\alpha}}

\def\lya{Ly$\alpha$}

\def\pcite#1{(\cite{#1})}
\def\pref#1{(\ref{#1})}

\def\macho{{\sc macho}}
\def\newpage{\vfill\eject}
\def\vs{\vskip 0.2truein}
\def\gnu{\Gamma_\nu}
\def\fnu {{\cal F_\nu}}
\def\mass{m}
\def\lum{{\cal L}}
\def\imf{\Psi(\mass)}
\def\ilf{\Phi(M)}
\def\msun{M_\odot}
\def\zsun{Z_\odot}
\def\met{[M/H]}
\def\vi{(V-I)}
\def\mtot{M_{\rm tot}}
\def\mhalo{M_{\rm halo}}
\def\pp{\parshape 2 0.0truecm 16.25truecm 2truecm 14.25truecm}
\def\la{\mathrel{\mathpalette\fun <}}
\def\ga{\mathrel{\mathpalette\fun >}}
\def\fun#1#2{\lower3.6pt\vbox{\baselineskip0pt\lineskip.9pt
  \ialign{$\mathsurround=0pt#1\hfil##\hfil$\crcr#2\crcr\sim\crcr}}}
\def\ie{{\it i.e., }}
\def\eg{{\it e.g., }}
\def\etal{{\it et al. }}
\def\etalc{{\it et al., }}
\def\kpc{{\rm kpc}}
 \def\Mpc{{\rm Mpc}}
\def\mh{\mass_{\rm H}}
\def\mmax{\mass_{\rm u}}
\def\ml{\mass_{\rm l}}
\def\bc{f_{\rm cmpct}}
\def\br{f_{\rm rd}}
\def\kmsec{{\rm km/sec}}
\def\ibl{{\cal I}(b,l)}
\def\dmax{d_{\rm max}}

\title{LIMITS ON STELLAR OBJECTS AS THE DARK MATTER OF OUR HALO:
NONBARYONIC DARK MATTER SEEMS TO BE REQUIRED}
\author{Katherine Freese$^{1}$, Brian Fields$^2$, and David Graff$^3$}
\address{1)University of Michigan\\ The Harrison M. Randall Laboratory of
Physics\\ Ann Arbor, MI 48109-1120 USA \\ 2) University of Illinois\\
Dept. of Astronomy\\ Champagne-Urbana, IL USA \\ 3) Ohio State
University\\ Dept. of Astronomy\\ Columbus, OH USA\\
{\rm 1) ktfreese@umich.edu,
2)bdfields@astro.uiuc.edu, 3)graff.25@osu.edu}}



\begin{abstract}
The nature of the dark matter in the Halo of our Galaxy remains a
mystery.  Arguments are presented that the dark matter does not consist
of ordinary stellar or substellar objects, i.e., the dark matter is not
made of faint stars, brown dwarfs, white dwarfs, or neutron stars. In
fact, faint stars and brown dwarfs constitute no more than a few percent
of the mass of our Galaxy, and stellar remnants must satisfy
$\Omega_{\rm WD} \leq 3 \times 10^{-3} h^{-1}$, where $h$ is the Hubble
constant in units of 100 km s$^{-1}$ Mpc$^{-1}$.  On theoretical grounds
one is then pushed to more exotic explanations.  Indeed a nonbaryonic
component in the Halo seems to be required.
\end{abstract}

\section{Introduction}
\setcounter{footnote}{0}
\renewcommand{\thefootnote}{\arabic{footnote}}

The nature of the dark matter in the haloes of galaxies is an
outstanding problem in astrophysics.  Over the last several decades
there has been great debate about whether this matter
is baryonic or must be exotic.  Many astronomers believed
that a stellar or substellar solution to this problem might be
the most simple and therefore most plausible explanation.
However, in the last few years, these candidates have been
ruled out as significant components of the Galactic Halo.
I will discuss limits on these stellar candidates, and
argue for my personal conviction that:
{\bf Most of the dark matter in the Galactic Halo must be nonbaryonic.}

Until recently, stellar candidates for the dark matter, including faint
stars, brown dwarfs, white dwarfs, and neutron stars, were extremely
popular.  However, recent analysis of various data sets has shown that
faint stars and brown dwarfs probably constitute no more than a few
percent of the mass of our Galaxy (Bahcall, Flynn, Gould, and Kirhakos
\cite {bfgk}); Graff and Freese \cite{gf96a}; Graff and Freese
\cite{gf96b}; Mera, Chabrier and Schaeffer \cite{mcs}; Flynn, Gould, and
Bahcall \cite{fgb}; and Freese, Fields, and Graff \cite{freese}).
Specifically, using Hubble Space Telescope and parallax data (with some
caveats mentioned in the text), we showed that faint stars and brown
dwarfs contribute no more than 1\% of the mass density of the Galaxy.
Microlensing experiments, which were designed to look for Massive
Compact Halo Objects (MACHOs), also failed to find these light stellar
objects and place strong limits on dark matter candidates in the
$(10^{-7} - 10^{-2}) M_\odot$ mass range.

Recently white dwarfs have received attention as possible dark matter
candidates.  Interest in white dwarfs has been motivated by microlensing
events interpreted as being in the Halo, with a best fit mass of $\sim
0.5 \msun$.  However, I will show that stellar remnants including white
dwarfs and neutron stars are extremely problematic as dark matter
candidates, due to a combination of mass budget issues and chemical
abundances (Fields, Freese, and Graff 1998): A significant fraction of
the baryons of the universe would have to be cycled through the white
dwarfs (or neutron stars) and their main sequence progenitors; however,
in the process, an overabundance of carbon and nitrogen is produced, far
in excess of what is observed both inside the Galaxy and in the
intergalactic medium. Agreement with measurements of these elements in
the Ly$\alpha$ forest would require $\Omega_{\rm WD} h \leq 2 \times
10^{-4}$. Throughout, $h$ is the Hubble constant in units of 100 km
s$^{-1}$ Mpc$^{-1}$.  Some uncertainty in the yields of C and N from low
metallicity stars motivated us (Fields, Freese, and Graff 1999) to look
also at D and He$^4$, whose yields are far better understood.  The
abundances of D and He$^4$ can be kept in agreement with observations
only for low mass white dwarf progenitors $(m_{prog} \sim 2 M_\odot)$
and $\Omega_{\rm WD} < 0.003$.  In addition, another constraint arises
from considering the contribution of white dwarf progenitors to the
infrared background.  If galactic halos contain stellar remnants, the
infra-red flux from the remnant progenitors would contribute to the
opacity of multi-TeV $\gamma$-rays.  But the HEGRA experiment does see
multi-TeV $\gamma$-rays from the blazar Mkn501 at z= 0.034.  By
requiring that the optical depth due to $\gamma \gamma \rightarrow e^+
e^-$ be less than 1 for a source at z=0.034, we limit the cosmological
density of stellar remnants (Graff, Freese, Walker, and Pinsonneault
1999), $\Omega_{\rm WD} \leq (1-3) \times 10^{-3} h^{-1}$.  Hence white
dwarfs, brown dwarfs, faint stars, and neutron stars are either ruled
out or extremely problematic as dark matter candidates.

Then the puzzle remains: What are the 14 MACHO events that have been
interpreted as being in the Halo of the Galaxy?  Are some of them
actually located elsewhere, such as in the LMC itself? These questions
are currently unanswered.  As regards the dark matter in the Halo of our
Galaxy, one is driven to nonbaryonic constituents as the bulk of the
matter. Possibilities include supersymmetric particles, axions,
primordial black holes, or other exotic candidates.

\subsection{\bf Microlensing Experiments}

The MACHO (Alcock \etal \cite{macho:1yr}, \cite{macho:2yr}) and EROS
(Ansari \etal \cite{ansari}) experiments have attempted to find the dark
matter of our Galactic Halo by monitoring millions of stars in the
neighboring Large Magellanic Cloud (LMC), which is approximately
$(45-60)$ kpc away; they have monitored stars in the Small Magellenic
Cloud (SMC) as well.  When a Macho crosses the line of sight between a
star in the LMC and us, the Macho's gravity magnifies the light of the
background star.  The background star gets temporarily brighter and then
dims back down.  The Macho acts as a lens for the background star.  The
duration of the event scales as $\Delta t \propto {\sqrt{m} \over v}$,
where $m$ is the mass of the Macho and $v$ is the velocity perpendicular
to the line of sight.  Thus there is a degeneracy in the interpretation
of the data between $m$ and $v$.  To break the degeneracy, one has to
assume a galactic model, e.g., one has to assume that the lenses are in
the Halo of our Galaxy.  The three events in the first year MACHO data
had a typical timescale of 40 days, which corresponds (with the above
assumption) to a best fit mass for the Machos of $\sim 0.1\msun$.  With
reanalysis and more data, four years of data yield 14 events of longer
duration, 35-150 days (T. Axelrod \cite{axelrod}; this is the Einstein
diameter crossing time).  Thus the new best fit mass is roughly
$$m \sim 0.5 \msun \, . $$

From the experiments, one can estimate what fraction of the Halo is
made of Machos.  Using isothermal sphere models for the Galaxy with the
two year data, the Macho group estimated that 50\% (+30\%,-20\%) of the
Halo could be made of Machos.  However, this estimate depends
sensitively on the model used for the Galaxy. Gates, Gyuk, and Turner
\cite{ggt} ran millions of models and found that the number of models
vs. Halo mass fraction peaks at Machos comprising (0-30)\% of the Halo,
with virtually no models compatible with a 100\% Macho Halo.

Hence there is evidence that a {\bf nonbaryonic} component to the Halo
of our Galaxy is required.  Microlensing experiments have ruled out a
large class of possible baryonic dark matter components.  Substellar
objects in the mass range $10^{-7}\msun$ all the way up to
$10^{-2}\msun$ are ruled out by the experiments.  In this talk I will
discuss the heavier possibilities in the range $10^{-2}\msun$ to few
$\msun$.

\section{\bf Baryonic Candidates}

In this talk I will concentrate on baryonic candidates.  Hegyi and Olive
\cite{ho} ruled out large classes of baryonic candidates.  See also the
work of Carr \cite{carr}.  Until recently the most plausible remaining
possibilities for baryonic dark matter were 

\noindent
--Red Dwarfs ($0.2 \msun >$ mass $> 0.09 \msun$).  These are stars just
massive enough to burn hydrogen; they shine due to fusion taking place
in the core of the star. Thus these are very faint stars.

\noindent
--Brown Dwarfs (mass $< 0.09 \msun$).  These are sub-stellar
objects that cannot burn hydrogen.  They are too light to have
fusion take place in the interior. 

\noindent
--White Dwarfs (mass $\sim 0.6 \msun$).  These are the end-products
of stellar evolution for stars of mass $< 8 \msun$.

In this talk, I will present limits on red dwarfs (Graff and Freese
1996a), brown dwarfs (Graff and Freese 1996b), and white dwarfs (Graff,
Laughlin, and Freese \cite{glf}; Fields, Freese, and Graff \cite{ffg};
Fields, Freese, and Graff \cite{ffg2}; Graff, Freese, Walker, and
Pinsonneault \cite{gfwp}) as candidates for baryonic dark matter.

\section{Faint Stars and Brown Dwarfs}

The number of stellar objects grows with decreasing stellar
mass;. Hence, until recently, there was speculation that there might be
a large number of faint stars or brown dwarfs that are just too dim to
have been seen.  However, as I will argue these candidates (modulo
caveats below) have now been ruled out as dark matter candidates.  Faint
stars and brown dwarfs constitute no more than a few percent of the mass
of our Galactic Halo.

\subsection{\bf Faint Stars}

First we used Hubble Space Telescope data (Bahcall, Flynn, Gould, and
Kirhakos 1994) to limit the mass density in red dwarfs to less than 1\%
of the Halo (Graff and Freese 1996a).  The data of Bahcall et al (1994)
from HST examined a small deep field and measured the relative
magnitudes of stars in the V and I bands.
We used the six stars that were seen with $1.7 < V-I < 3$ to limit the
density of red dwarfs in the Halo.  First we obtained the distances to
these stars, which are shown in Figure 1.  One can see that the survey
is sensitive out to at least 10 kpc.  Note that the closest stars are
likely disk contaminants and not included in our final analysis.  We
obtained estimates of the stellar masses of these objects from stellar
models of Baraffe et al (1996); the masses are in the range
0.0875$\msun$ - 0.2$\msun$.

\begin{figure}
\centering
\psfig{figure=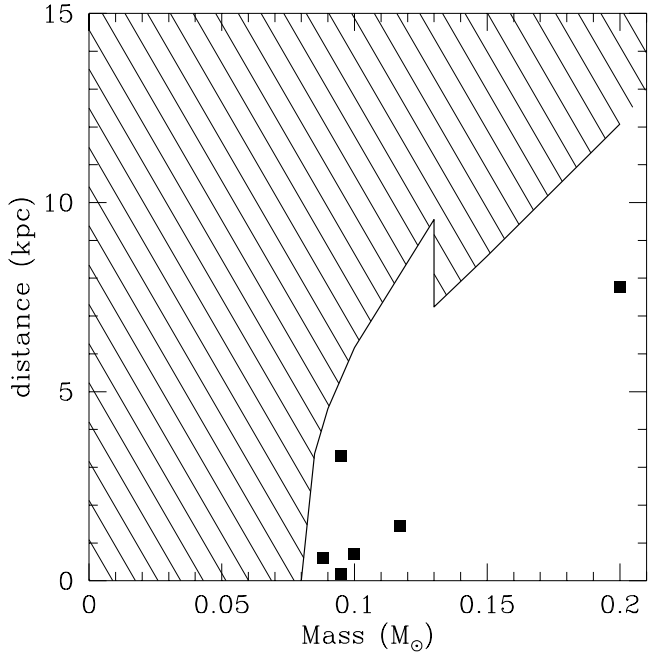, width=9cm}
\caption{(taken from Graff and Freese 1996a):
Distances to six stars in HST data with $1.7 < V-I < 3$
obtained by comparing apparent with absolute magnitudes
of these stars.}
\end{figure}

For the 6 stars in the HST data with $1.7 < V-I < 3$, we thus obtained a
Halo red dwarf mass density.  We then compared this red dwarf mass
density with virial estimates of the Halo density to see what fraction
is composed of red dwarfs.  We took a local Halo mass density of $\rho_o
\sim 9 \times 10^{-3} \msun/pc^3$.  Bahcall et al (1994) had made this
comparison by assuming that the red dwarfs had properties of stars at
the edge of the high metallicity main sequence; these authors found that
red dwarfs contribute less than 6\% of the Halo density.  However, Halo
red dwarfs are low metallicity objects, and we were thus motivated to
redo the analysis as outlined above. A ground-based search for halo red
dwarfs by Boeshaar, Tyson, and Bernstein (1994) found a much smaller
number.  We felt that a careful reinterpretation of the Bahcall et al
(1994) data was in order.  Our result is that Red dwarfs with $1.7 < V-I
< 3$ (i.e., mass 0.0875 $< M/\msun < 0.2)$, make up less than 1\% of the
Halo; our best guess is that they make up 0.14\% - 0.37\% of the mass of
the halo.  Subsequent examination of the Hubble Deep Field by Flynn,
Gould, and Bahcall \cite{fgb} and work by Mera, Chabrier, and Schaeffer
\cite{mcs2} reiterated that low-mass stars represent a negligible
fraction of the Halo dark matter.

\subsection{\bf Brown Dwarfs}

With these strong limits on the contribution of faint stars to the
Galactic Halo, we then obtained a Mass Function of these same red dwarfs
in order to be able to extrapolate to the brown dwarf regime; in this
way we were able to limit the contribution of brown dwarfs as well.  We
obtained the mass function from the following relation:
\begin{equation}
\label{derivemassfunc}
{\rm Mass Function} = {({\rm dM}_{\rm V} / {\rm dm})} \times  {\rm
Luminosity Function} \, .
\end{equation}
Here, the Mass Function (hereafter MF) is the number
density of stars with mass between $m$ and $m+dm$,
and the Luminosity Function (hereafter LF) is the number density of
stars in a magnitude range $M \rightarrow M+dM$ (note that $M$
refers to magnitude while $m$ refers to mass).
The luminosity function is what is observed; we used parallax
data taken by the US Naval Observatory (Dahn et al 1995) who
identified 114 halo stars.
We went from this observed luminosity function to the desired
mass function via stellar models of $M_V(m)$ obtained
by Alexander et al. \cite{models96}.

\begin{figure}
\centering 
\psfig{figure=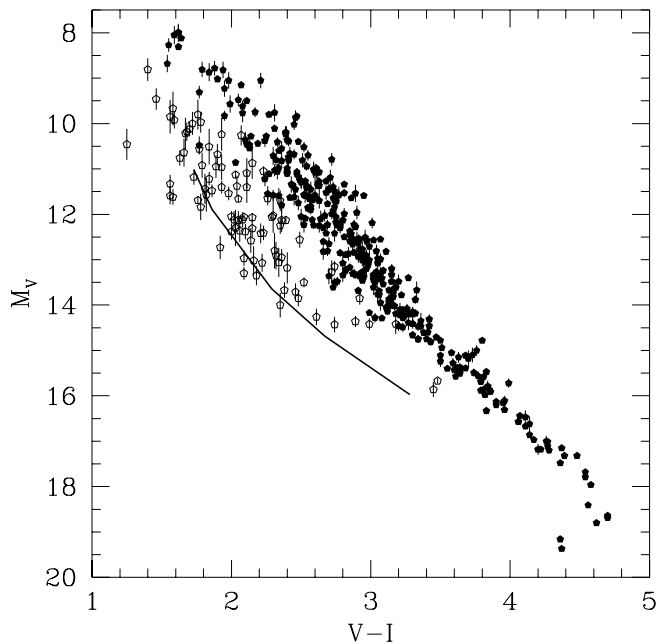, width=9cm} 
\caption{(taken from Dahn et al 1995): H-R diagram of nearby stars with measured
parallax.  The filled circles are high metallicity disk stars; the
velocity dispersion of these disk stars is $\sim 30$ km/sec.  The open
circles are low metallicity halo ``subdwarfs"; these stars have high
proper motions $\sim 200$ km/sec.  We have superimposed a solid line
which indicates the theoretical model of Baraffe et al (1995) with
log(Z/Z$_\odot) = -1.5$.}
\end{figure}
 
The parallax data (Dahn et al 1995) are shown in Figure 2. This is an
H-R diagram of nearby stars with measured parallax.  The filled circles
are high metallicity disk stars. The open circles are red dwarfs which
are known to be in the Halo because of their low metallicities and high
velocities.  It is these 114 Halo stars that we used to get a mass
function.  We always took the most ``conservative" case, i.e., the
steepest MF towards low mass; this case would give the largest number of
brown dwarfs and low mass red dwarfs.  For this reason, we considered a
number of metallicities and used the lowest realistic value of $Z = 3
\times 10^{-4}$.  There is a potential complication in that some of the
stars in the survey may actually be unresolved binaries.  If so, the
observed light is the sum of the light from two stars.  Then one may
overestimate the mass of the star if one assumes the light is from a
single star.  We considered three models for binaries.  The most extreme
of these is that all the stars are really in binaries, with equal masses
for the two stars in the binary system.  Then the luminosity of each
star is really half as big as if it had been a single star, each star
has a smaller mass, and one obtains a steeper mass function towards low
mass.  This model is unphysical but simple, and we used it to illustrate
an extreme for the largest number of stars at low mass that can be
obtained from this data set.  Figure 3 shows the mass functions that we
obtained, for the case of no binaries and the extreme case of 100\%
binaries.  In these plots we multiplied the vertical axis by $m^2$ for
simplicity of interpretation.  With this factor of $m^2$, a mass
function (MF) that is decreasing to the left converges, an MF that is
increasing to the left diverges, while an MF that is flat diverges only
logarithmically.  In figure 3a, the case of no binaries, we can see that
the MF $\times m^2$ decreases to the left (convergent); in Figure 4c,
the case of 100\% binaries, the MF $\times m^2$ is flat (diverges
logarithmically).  Hence Figure 3 summarizes our results for the mass
function for faint stars heavier than $0.09\msun$.

\begin{figure}
\centering
\psfig{figure=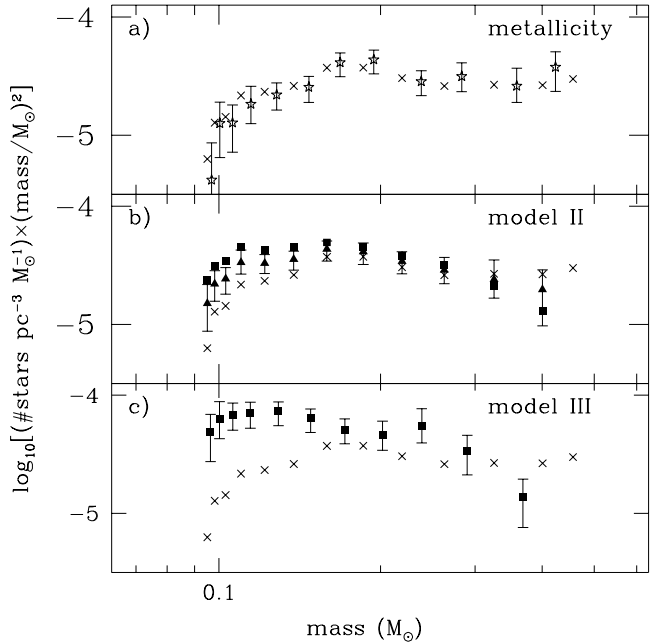, width=9cm}
\caption{(taken from Graff and Freese 1996b): The mass
function of red dwarf halo stars (multiplied by m$^2$).  Each of the
four models is derived from the LF of Dahn ${\it et \, al}$ (1995) but
assumes different metallicity and binary content.  In all three panels,
crosses without errorbars illustrate the mass function derived for stars
with metallicity Z = $3 \times 10^{-4}$ and no binary companions.  The
other model presented in panel (a) has Z = $6 \times 10^{-4}$ (no
binaries) for comparison.  Panels (b) and (c) show binary models II and
III for $Z = 3 \times 10^{-4}$.  Binary model III has been designed to
exaggerate the number of low mass stars compared to high mass ones and
is unrealistic.}
\end{figure}

Now, in order to proceed with an extrapolation of this red dwarf mass
function past the hydrogen burning limit into the red dwarf regime, we
need a brief theoretical interlude.  Star formation theory indicates
that, as one goes to lower masses, the MF rises no faster than a power
law.  The theories of Adams and Fatuzzo (1996), Larson (1992), Zinnecker
(1984), and Price and Podsiadlowski (1995), while based on different
physical principles, all find this same upper limit.  Hence we looked
for the power law describing the red dwarf mass function at the lowest
masses, and then use this same power law to extrapolate into the brown
dwarf regime.  We took the mass function to scale as
\begin{equation}
\label{massfunc}
MF\propto m^{- \alpha} \, .
\end{equation}
Then the total mass in the Halo is
\begin{equation}
\label{mtot}
m_{tot} = \int_0^{0.09\msun} m \times MF \times dm \, .
\end{equation}
If $\alpha > 2$, then the total mass diverges.  If $\alpha = 2$,
then the total mass diverges only logarithmically.  If $\alpha <2$,
then the total mass converges.  We found
\begin{equation}
\label{alpha}
\alpha \leq 2 \, ,
\end{equation}
for all models.  More specifically, for the extreme case
of 100\% binaries, we found $\alpha = 2$, i.e., each order of
magnitude of mass range contains an equal total mass.  Even for
a lower limit $\sim m_{moon}$, the total mass in brown dwarfs
is less than 3\% of the Halo mass.  For all other models,
including the case of no binaries, we find $\alpha <2$,
and brown dwarfs constitute less than a percent of the Halo mass.
Similar results were found by Mera, Chabrier, and Schaeffer \cite{mcs}.

How might one avoid these conclusions?  First, star formation theory
might be completely wrong.  Alternatively, there might be a spatially
varying initial mass function so that brown dwarfs exist only at large
radii and not in our locality, so that they were missed in the data
(Kerins and Evans \cite{kerev}).

The two year MACHO microlensing data have also shown that, for standard
Halo models as well as a wide range of alternate models, the timescales
fo the events are not compatible with a population of stars lighter than
0.1$\msun$ (Gyuk, Evans, and Gates 1998).

\subsection{\bf Punchline}

The basic result of this work is that the total mass density of local
Population II Red Dwarfs and Brown Dwarfs makes up less than 1\% of the
local mass density of the Halo; in fact, these objects probably make up
less than $0.3\%$ of the Halo.

\section{Mass Budget Issues}

This section (based on work by Fields, Freese, and Graff \cite{ffg}) is
general to all Halo Machos, no matter what kind of objects they are.

\subsection{\bf Contribution of Machos to the Mass Density of the
Universe}

Dalcanton \etal were able to place strong limits on the cosmological
mass density of Machos even before the galactic microlensing experiments
produced their first results.  They looked for a reduction in apparent
equivalent width of quasar emission lines; such a reduction would be
caused by compact objects.  They found that $\Omega_m < 0.1$.

There is a potential problem in that too many baryons are tied up in
Machos and their progenitors (Fields, Freese, and Graff \cite{ffg}).  We
begin by
estimating the contribution of Machos to the mass density of the
universe: Microlensing results (Alcock \etal 1997a) predict that the
total mass of Machos in the Galactic Halo out to 50 kpc is
\begin{equation}
\label{outto}
M_{\rm Macho} = (1.3 - 3.2) \times 10^{11} \msun \, .
\end{equation}
Now one can obtain a ``Macho-to-light" ratio for the Halo by
dividing by the luminosity of the Milky Way (in the B-band),
\begin{equation}
\label{lummw}
L_{MW} \sim (1.3-2.5) \times 10^{10} L_\odot \, .
\end{equation}
We obtain
\begin{equation}
\label{masstolight}
(M/L)_{\rm Macho} = (5.2-25)\msun/L_\odot \, .
\end{equation}
From the ESO Slice Project
Redshift survey (Zucca \etal \cite{zuc}),
the luminosity
density of the Universe in the $B$ band is
\begin{equation}
\label{phi}
{\cal L}_B = 1.9\times 10^{8} h \ L_\odot \ {\rm Mpc}^{-3}
\end{equation}
where the Hubble parameter
$h=H_0/(100 \, {\rm km} \, {\rm sec}^{-1} \, \Mpc^{-1})$.
If we assume that the $M/L$ which we defined for the Milky
Way is typical of the Universe as a whole,
then the universal mass density of
Machos is
\begin{equation}
\label{rho}
\rho_{\rm Macho}
   = (M/L)_{\rm Macho} \, {\cal L}_B
   = (1-5) \times 10^{9} h \ \msun \, {\rm Mpc}^{-3} \, .
\end{equation}
The corresponding fraction of the critical density
$\rho_c \equiv
3H_0^2/8 \pi G = 2.71 \times 10^{11} \, h^2 \, M_\odot \ \Mpc^{-3}$ is
\begin{equation}
\label{omega}
\Omega_{\rm Macho} \equiv \rho_{\rm Macho}/ \rho_c = (0.0036-0.017) \,
h^{-1} \, .
\end{equation}
Note: see also the discussion by Fukugita, Hogan, and Peebles (\cite{fhp}).

We will now proceed to compare our $\omegam$ derived in Eq.\
\pref{omega} with the baryonic density in the universe, $\omegab$, as
determined by primordial nucleosynthesis.  Recently, the status of Big
Bang nucleosynthesis has been the subject of intense discussion,
prompted both by observations of deuterium in high-redshift quasar
absorption systems, and also by a more careful examination of
consistency and uncertainties in the theory.  To conservatively allow
for the full range of possibilities, we will therefore adopt
\begin{equation}
\label{omegab}
\omegab= (0.005-0.022) \ h^{-2} \, .
\end{equation}

We can see that $\omegam$ and $\omegab$ are roughly comparable within
this na\"i ve calculation.  Thus, if the Galactic halo Macho
interpretation of the microlensing results is correct, Machos make up an
important fraction of the baryonic matter of the Universe.
Specifically, the central values in eqs.\ (\ref{omega}) and
(\ref{omegab}) give
\begin{equation}
\label{central}
\omegam/\omegab \sim 0.7 \, .
\end{equation}
However, the lower limit on this fraction is
considerably smaller and hence less restrictive.
Taking the lowest possible
value for $\omegam$ and the highest possible value for $\omegab$,
we see that
\begin{equation}
\label{comp}
{\omegam \over \omegab} \geq {1 \over 6} h \geq \frac{1}{12} \, .
\end{equation}

The only way to avoid these conclusions is to argue that the luminosity
density in eqn. (8) is dominated by galaxies without Machos, so that the
Milky Way is atypically rich in Machos.  However, this is extremely
unlikely, because most of the light contributing to the luminosity
density ${\cal{L}}$ comes from galaxies similar to ours.  Even if Machos
only exist in spiral galaxies (2/3 of the galaxies) within one magnitude
of the Milky Way, the value of $\Omega_{\rm Macho}$ is lowered by at
most a factor of 0.17.

\subsection{\bf Comparison with the Lyman-${\bf \alpha}$ Forest}

We can compare the Macho contribution to other components of the
baryonic matter of the universe.  In particular, measurements of the
Lyman-$\alpha$ (\lya) forest absorption from intervening gas in the
lines of sight to high-redshift QSOs indicate that many, if not most, of
the baryons of the universe were in this forest at redshifts $z>$2.  It
is hard to reconcile the large baryonic abundance estimated for the
\lya\ forest with $\omegam$ obtained previously (Gates, Gyuk, Holder, \&
Turner \cite{gght}).  Although measurements of the \lya\ forest only
obtain the neutral column density, careful estimates of the ionizing
radiation can be made to obtain rough values for the total baryonic
matter, i.e. the sum of the neutral and ionized components, in the \lya\
forest.  For the sum of these two components, Weinberg et
al. \pcite{wmhk} estimate
\begin{equation}
\label{lyman}
\Omega_{\rm Ly\alpha} \sim 0.02 h^{-3/2} \, .
\end{equation}
This number is at present uncertain.  For example, it assumes an
understanding of the UV background responsible for ionizing the IGM, and
accurate determination of the quasar flux decrement due to the neutral
hydrogen absorbers.  Despite these uncertainties, we will use Eq.\
({\ref{lyman}}) below and examine the implications of this estimate.

We can now require that the sum of the Macho energy density
plus the \lya\ baryonic energy density do not
add up to a value in excess of the baryonic density from
nucleosynthesis:
\begin{equation}
\label{insist}
\omegam(z) + \omegalya (z) \leq \omegab \, ;
\end{equation}
this expression holds for any epoch $z$.  Unfortunately, the
observations of Machos and \lya\ systems are available for different
epochs.  Thus, to compare the two one must assume that there has not
been a tradeoff of gas into Machos between the era of the Lyman systems
($z \sim 2-3$) and the observation of the Machos at $z=0$.  That is, we
assume that the Machos were formed before the \lya\ systems.

Although Eq.\ (\ref{insist}) offers a potentially strong constraint, in
practice the uncertainties in both $\omegalya$ and in $\omegab$ make a
quantitative comparison difficult.  Nevertheless, we will tentatively
use the numbers indicated above.  We then have
\begin{equation}
\label{onehalf}
(\omegam = 0.007-0.04) + (\omegalya = 0.06) \leq (\omegab = 0.02 - 0.09)
 \ {\rm for} \, h=1/2 \, ,
\end{equation}
and
\begin{equation}
\label{one}
(\omegam = 0.004 - 0.02) + (\omegalya = 0.02) \leq (\omegab = 0.005 - 0.02)
\ {\rm for} \, h=1 \, .
\end{equation}
These equations can be satisfied, but
only if one uses the most favorable extremes
in both $\omegam$ and $\omegab$, i.e., for
the lowest possible values for $\omegam$
and the highest possible values for $\omegab$.

Recent measurements of Kirkman and Tytler \pcite{kt} of the ionized
component of a Lyman limit system at z=3.3816 towards QSO HS 1422+2309
estimate an even larger value for the mass density in hot and highly
ionized gas in the intergalactic medium: $\Omega_{\rm hot} \sim 10^{-2}
h^{-1}$.  If this estimate is correct, then Eq.\ (\ref{insist}) becomes
even more difficult to satisfy.

One way to avoid this mass budget problem would be to argue that the
\lya\ baryons later became Machos.  Then it would be inappropriate to
add the \lya\ plus Macho contributions in comparing with $\omegab$,
since the Machos would be just part of the \lya\ baryons.  However, the
only way to do this would be to make the Machos at a redshift after the
\lya\ measurements were made.  Since these measurements extend down to
about $z\sim 2-3$, the Machos would have to be made at $z<2$.  However,
this would be difficult to maneuver.  A large, previously unknown
population of stellar remnants could not have formed after redshift 2;
we would see the light from the stars in galaxy counts (Charlot and Silk
\cite{cs}) and in the Hubble Deep Field (Loeb \cite{loeb}).

Until now we have only considered the contribution to the baryonic
abundance from the Machos themselves.  Below we will consider the
baryonic abundance of the progenitor stars as well, in the case where
the Machos are stellar remnants.  When the progenitor baryons are added
to the left hand side of Eq.\ (\ref{insist}), this equation becomes
harder to satisfy.  However, we wish to reiterate that measurements of
$\Omega_{{\rm Ly} \alpha}$ are at present uncertain, so that it is
possibly premature to conclude that Machos are at odds with the amount
of baryons in the Ly$\alpha$ forest.

\section{Machos as Stellar Remnants: White Dwarfs or Neutron Stars}

In the last section on the mass budget of Machos, we assumed merely
that they were baryonic compact objects.  In this section
(based on work by Fields, Freese, and Graff \cite{ffg}, Fields, Freese,
and Graff \cite{ffg2}, and Graff, Freese, Walker, and Pinsonneault
\cite{gfwp}):
we turn to the specific possibility that Machos are stellar remnants
white dwarfs, neutron stars, or black holes.
The most complete microlensing data indicate a best fit mass
for the Machos of roughly $(0.1 - 1)\msun$.  Hence there has
been particular interest in the possibility that these objects
are white dwarfs.  I will discuss problems and issues
with this interpretation: in particular I will discuss the
baryonic mass budget and the pollution due to white dwarf
progenitors.

\subsection{\bf Mass Budget Constraints from the Macho Progenitors}

In general, white dwarfs, neutron stars, or black holes all came from
significantly heavier progenitors.  Hence, the excess mass left over
from the progenitors must be added to the calculation of $\Omega_{\rm
Macho}$; the excess mass then leads to stronger constraints.  Previously
we found that any baryonic Machos that are responsible for the Halo
microlensing events must constitute a significant fraction of all the
baryons in the universe.  Here we show that, if the Machos are white
dwarfs or neutron stars, their progenitors, while on the main sequence,
are an even larger fraction of the total baryonic content of the
universe.  The excess mass is then ejected in the form of gas when the
progenitors leave the main sequence and become stellar remnants.  This
excess mass is quite problematic, as there is more of it than is allowed
by big band nucleosynthesis and it is chemically enriched beyond what is
allowed by observations of Halo stars and the intergalactic medium.

If all the Machos formed in $< 1$Gyr (the burst model), then
(for different choices of the initial mass function)
we can determine the additional contribution of the
excess gas to the mass density of the universe.  Typically
we find the contribution of Macho progenitors to
the mass density of the universe to be
\begin{equation}
\label{omegaprog}
\Omega_{{\rm prog}} = 4 \Omega_{{\rm Macho}} = (0.016-0.08)h^{-1} \, .
\end{equation}
(As an extreme minimum, we find an enhancement factor of 2 rather than
4).  From comparison with $\omegab$, we can see that a very large
fraction of the baryons of the universe must be cycled through the
Machos and their progenitors.  In fact, the central values of all the
numbers now imply
\begin{equation}
\label{toomuch}
\Omega_{\rm prog} \sim 3 \Omega_B \, ,
\end{equation}
which is obviously unacceptable.  One is driven to the lowest values of
$\Omega_{rm Macho}$ and highest value of $\Omega_B$ to avoid this
problem.

\subsection{\bf Galactic Winds}

The white dwarf progenitor stars return most of their mass in their
ejecta, i.e., planetary nebulae composed of processed material.  Both
the mass and the composition of the material are potential problems.  As
we have emphasized, the cosmic Macho mass budget is a serious issue.
Here we see that it is significant even when one considers only the
Milky Way.  The amount of mass ejected by the progenitors is far in
excess of what can be accommodated by the Galaxy.  Given the $M_{\rm
Macho}$ of Eq.(5), a burst model requires the total mass of progenitors
in the Galactic Halo (out to 50 kpc) to have been at least twice the
total mass in remnant white dwarfs, i.e., $M_{{\rm prog}} \ge 2 M_{\rm
Macho} = (2.4 - 5.8) \times 10^{11} \msun$.  The gas that is ejected by
the Macho progenitors is collisional and tends to fall into the Disk of
the Galaxy.  But the mass of the ejected gas $M_{\rm gas} = M_{{\rm
prog}} - M_{\rm Macho} \sim M_{\rm Macho}$ is at least as large as the
mass ($\sim 10^{11} M_\odot$) of the Disk and Spheroid of the Milky Way
combined.  We see that the Galaxy's baryonic mass budget---including
Machos---immediately demands that some of the ejecta be {\it removed}
from the Galaxy.

This requirement for outflow is intensified when one considers the
composition of the stellar ejecta.  It will be void of deuterium, and
will include large amounts of the nucleosynthesis products of $(1-8)
\msun$ white dwarf progenitors, notably: helium, carbon, and nitrogen
(and possibly s-process material).

A possible means of removing these excess baryons is a Galactic wind.
Indeed, as pointed out by Fields, Mathews, \& Schramm \pcite{fms}, such
a wind may be a virtue, as hot gas containing metals is ubiquitous in
the universe, seen in galaxy clusters and groups, and present as an
ionized intergalactic medium that dominates the observed neutral \lya\
forest.  Thus, it seems mandatory that many galaxies do manage to shed
hot, processed material.

Such a wind may be driven by some of the white dwarfs themselves
(Fields, Freese, and Graff \cite {ffg2}).  Some of the white dwarfs may
accrete from binary red giant companions and give rise to Type I
Supernovae, which serve as an energy source for Galactic winds.
However, excess heavy elements such as Fe are overproduced in the
process (Canal, Isern, and Ruiz--Lapuente \cite{pilar}).

\subsection{\bf On Carbon and Nitrogen}
\label{sect:carbon}

The issue of carbon (Gibson \& Mould \cite{gm}) and/or nitrogen produced
by white dwarf progenitors is the greatest difficulty faced by a white
dwarf dark matter scenario.  Stellar carbon yields for zero metallicity
stars are quite uncertain.  Still, according to the Van den Hoek \&
Groenewegen (1997) yields, a star of mass 2.5$\msun$ will produce about
twice the solar enrichment of carbon.  If a substantial fraction of all
baryons pass through intermediate mass stars, the carbon abundance in
this model will be near solar.

Then overproduction of carbon can be a serious problem, as emphasized by
Gibson \& Mould \pcite{gm}.  They noted that stars in our galactic halo
have carbon abundance in the range $10^{-4}-10^{-2}$ solar, and argued
that the gas which formed these stars cannot have been polluted by the
ejecta of a large population of white dwarfs.  The galactic winds
discussed in the previous section could remove carbon from the star
forming regions and mix it throughout the universe.

However, carbon abundances in intermediate redshift \lya\ forest lines
have recently been measured to be quite low.  Carbon is indeed present,
but only at the $\sim 10^{-2}$ solar level, (Songaila \& Cowie
\cite{sc}) for \lya\ systems at $z \sim 3$ with column densities $N \ge
3 \times 10^{15} \, {\rm cm}^{-2}$.  Ly$\alpha$ forest abundances have
also been recently measured at low redshifts with HST (Shull \etal
\cite{shull}) to be less than $3 \times 10^{-2}$ solar.  Furthermore, in
an ensemble average of systems within the redshift interval $2.2 \le z
\le 3.6$, with lower column densities ($10^{13.5} \, {\rm cm}^{-2} \le N
\le 10^{14} \, {\rm cm}^{-2}$), the mean C/H drops to $\sim 10^{-3.5}$
solar (Lu, Sargent, Barlow, \& Rauch \cite{lsbr}).

In order to maintain carbon abundances as low as $10^{-2}$ solar, only about
$10^{-2}$ of all baryons can have passed through the intermediate mass
stars that were the predecessors of Machos.  Such a fraction can barely
be accommodated by our results in section 4.1
for the remnant density predicted from our extrapolation
of the Macho group results, and would be in conflict with
$\Omega_{{\rm prog}}$ in the case of a single burst of star formation.

We note that progenitor stars lighter than 4$\msun$ overproduce Carbon;
whereas progenitor stars heavier than 4$\msun$ may replace the carbon
overproduction problem with nitrogen overproduction (Fields, Freese,
and Graff \cite{ffg2}).   The heavier
stars may have a process known as Hot Bottom Burning, in which
the temperature at the bottom of the star's convective envelope
is high enough for nucleosynthesis to take place, and carbon
is processed to nitrogen (Lattanzio \cite{latt},
Renzini and Voli \cite {rv}, Van den Hoek and Groenewegen (1997),
Lattanzio and Boothroyd \cite{lattbooth}).
In this case one gets a ten times
solar enrichment of nitrogen, which is far in excess of
the observed nitrogen in damped Lyman systems.
In conclusion, both C and N exceed what's observed.

Using the yields described above, we calculated the C and N that would
result from the stellar processing for a variety of initial mass
functions for the white dwarf progenitors.  We used a chemical evolution
model based on a code described in Fields \& Olive \cite{fo98} to obtain
our numerical results.  The star formation rate is chosen as an
exponential $\psi \propto e^{-t/\tau}$ with an $e$-folding time $\tau =
0.1$ Gyr, although we have found that the results are insensitive to
details of the star formation rate up to $\tau = 1$ Gyr.  Our results
are presented in panels b) of figures 4 and 5.  The CN abundances are
presented relative to solar via the usual notation of the form
\begin{equation}
[{\rm C/H}]= \log_{10} \frac{{\rm C/H}}{({\rm C/H})_\odot} \, .
\end{equation}
For example, in this notation $[{\rm C/H}]=0$ represents
a solar abundance of C, while $[{\rm C/H}]=-1$is 1/10 solar.
Our C and N abundances were obtained without including HBB, which
would exchange a C overproduction problem for a N overproduction problem.

\begin{figure}
\centering
\psfig{figure=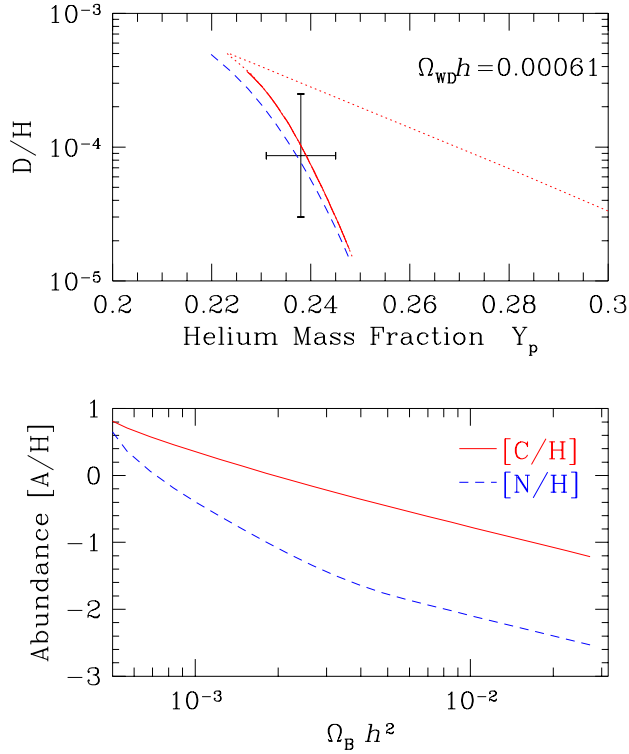, width=9cm}
\caption{(taken from Fields, Freese, and Graff 1999): {\bf
(a)} The D/H abundances and helium mass fraction $Y$ for models with
$\Omega_{\rm WD} h = 6.1 \times 10^{-4}$, $h=0.7$, and IMF peaked at
$2\msun$.  The red curves show the changes in primordial D and He and a
result of white dwarf production. The solid red curve is for the full
chemical evolution model, the dotted red curve is for instantaneous
recycling, and the long-dashed red curve for the burst model.  The
short-dashed blue curve shows the initial abundances; the error bars
show the range of D and He measurements.  This is the absolute minimum
$\Omega_{\rm WD}$ compatible with cosmic extrapolation of white dwarf
Machos if Machos are contained only in spiral galaxies with luminosities
similar to the Milky Way.  \hfill\break {\bf (b)} CNO abundances
produced in the same model as {\bf a}, here plotted as a function of
$\Omega_B$.  The C and N production in particular are greater than 1/10
solar.}
\end{figure}

In Figure 4, we make the parameter choices that are in agreement with D
and He$^4$ measurements (see the discussion below) and are the least
restrictive when comparing with the Ly$\alpha$ measurements.  We take
$\Omega_{\rm WD} h = 6.1 \times 10^{-4}$, the minimum amount of WD
required to explain the microlensing results if only galaxies similar to
ours produce WD Machos.  We take h= 0.7 and an initial mass function
(IMF) sharply peaked at 2$\msun$, so that there are very few progenitor
stars heavier than 3$\msun$ (this IMF is required by D and He$^4$
measurements).  In Figure 5, we have the same values for $\Omega_{\rm
WD}$ and h but have taken an IMF peaked at 4$\msun$.  In both cases, by
comparing with the observations, we obtain the limit,
\begin{equation}
\Omega_{\rm WD} h \leq 2 \times 10^{-4} \, .
\end{equation}
We also considered a variety of other parameter choices, and obtained
the same limit (see the figures in Fields, Freese, and Graff 1999).

\begin{figure}
\centering
\psfig{figure=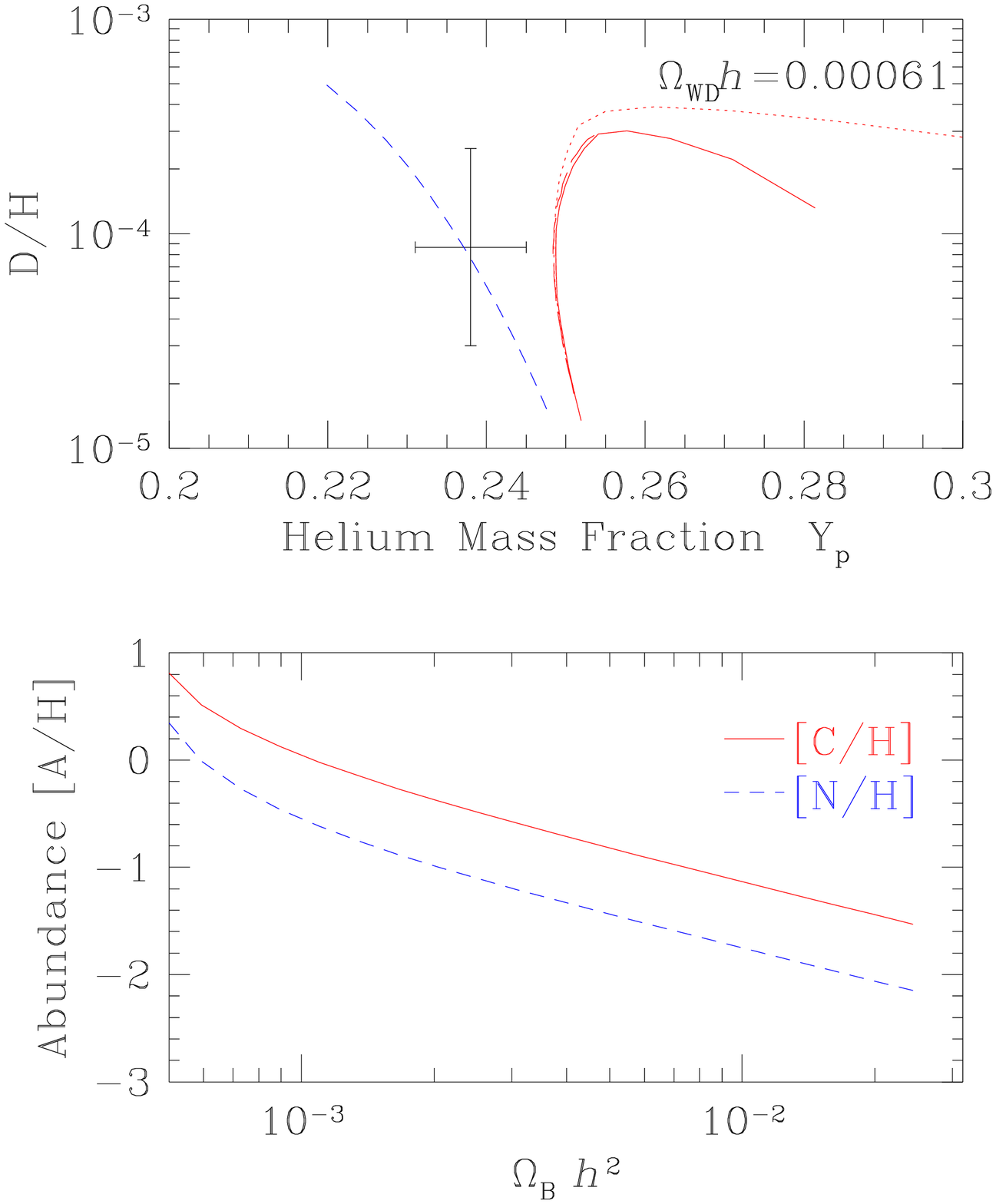, width=9cm}
\caption{As in Fig. 4, but with an IMF peaked at 4$\msun.$ We
see that the processing drives D and He out of the measured range.}
\end{figure}

Alternatively, we require an actual abundance distribution that
is quite heterogeneous: those regions in which the
observations are made must be underprocessed.  This implies segregation
efficiency of 97\%.

Note that it is possible (although not likely) that carbon never leaves
the white dwarf progenitors, so that carbon overproduction is not a
problem (Chabrier \cite{chabriernew}).  Carbon is produced exclusively
in the stellar core.  In order to be ejected, carbon must convect to the
outer layers in the ``dredge up'' process.  Since convection is less
efficient in a zero metallicity star, it is possible that no carbon
would be ejected in a primordial star.  In that case, it would be
impossible to place limits on the density of white dwarfs using carbon
abundances.  We have here assumed that carbon does leave the white dwarf
progenitor stars.

\subsection{\bf Deuterium and Helium}

Because of the uncertainty in the C and N yields from low-metallicity
stars, we have also calculated the D and He$^4$ abundances that
would be produced by white dwarf progenitors.  These are far less uncertain
as they are produced farther out from the center of the star and
do not have to be dredged up from the core.
We use both the numerical model discussed above, in which the stars
have finite lifetimes, and also two extreme analytical models to bracket
the possible results.  We consider a burst model, in which the timescale
for star formation is much shorter than the lifetimes of the stars.
We also consider the  opposite limit, the
instantaneous recycling approximation, in which
the stellar lifetime is short compared to the star formation timescale.
In the figures we present results from both analytical approaches
and from the numerical model; we can see that the numerical results
are closely approximated by the burst model.

Panels a) in Fig. 4  and 5 display our results.  Also shown are the initial
values from big bang nucleosynthesis and the (very generous) range of
primordial values of D and He$^4$ from observations.
One can see right away that Fig. 4 obtains abundances
compatible with the measurements, while
the model in Fig. 5 fails to match the D and He$^4$ measurements.
Thus, from D and He alone, we can see that the white dwarf progenitor IMF
must be peaked at low masses, $\sim 2\msun$.
We obtain
\begin{equation}
\Omega_{\rm WD} \leq 0.003 \, .
\end{equation}

\subsection{\bf Background Light}

If galactic halos contain stellar remnants, the infrared flux from
the remnant progenitors would contribute to the opacity of multi-TeV
$\gamma$-rays (Konopelko \etal \cite{konop}).
The multi-TeV $\gamma$-ray horizon is
established to be at a redshift $z>0.034$ by the observation
of the blazar Mkn501.  By requiring that the optical depth
due to $\gamma \gamma \rightarrow e^+ e^-$ be less than one
for a source at $z=0.034$ we limit the cosmological density
of stellar remnants (Graff, Freese, Walker, and Pinsonneault \cite{gfwp}),
\begin{equation}
\Omega_{\rm WD} \leq (1-3) \times 10^{-3}
h^{-1} \, .
\end{equation}
In other words, if the density of white dwarfs
exceeds this value, the infrared radiation from the progenitors
would have prevented TeV $\gamma$-rays from Mkn501 from ever reaching us.

\subsection{\bf Neutron Stars}

The first issue raised by neutron star Macho candidates is their
compatibility with the microlensing results.  Neutron stars ($\sim 1.5
\msun$) and stellar black holes ($\ga 1.5 \msun$) are more massive
objects, so that one would typically expect longer lensing timescales
than what is currently observed in the microlensing experiments (best
fit to $\sim 0.5 \msun$).  As discussed by Venkatesan, Olinto, \& Truran
\pcite{vot}, one must posit that as the experiments continue to take
measurements, longer timescale events should begin to be seen.  In this
regard, it is intriguing that the first SMC results
(Palanque-Delabrouille et al.\ \cite{eros:smc}; Alcock et al.\
\cite{macho:smc}) suggest lensing masses of order $\sim 2 \msun$.  Note
that these long timescales could be explained if the SMC events are due
to SMC self lensing (Palanque-Delabrouille \etal \cite{eros:smc}; Graff
\& Gardiner \cite{gg}).

However, the same issues of mass budget and chemical enrichment arise
for neutron stars as did for white dwarfs, only the problems are worse.
In particular, the higher mass progenitors of neutron stars eject even
more mass, so that $\Omega_{\rm prog}$ is even bigger than for the case
of white dwarfs.  The ejecta are highly metal rich and would need a
great deal of dilution (as much as for the case of white dwarfs) in
order to avoid conflict with observations. However, most of the baryons
in the universe have already been used to make the progenitors (even
more than for the case of white dwarfs); there are no baryons left over
to do the diluting.

\subsection{\bf Mass Budget Summary}

If Machos are indeed found in halos of galaxies like our own, we have
found that the cosmological mass budget for Machos requires $\omegam/
\omegab \geq {1 \over 6} h f_{\rm gal}$, where $f_{\rm gal}$ is the
fraction of galaxies that contain Machos, and quite possibly $\omegam
\approx \omegab$.  Specifically, the central values in eqs.\
(\ref{omega}) and (\ref{omegab}) give $\omegam/\omegab \sim 0.7$.  Thus
a stellar explanation of the microlensing events requires that a
significant fraction of baryons cycled through Machos and their
progenitors. If the Machos are white dwarfs that arose from a single
burst of star formation, we have found that the contribution of the
progenitors to the mass density of the universe is at least a factor of
two higher, probably more like three or four.  We have made a comparison
of $\omegab$ with the combined baryonic component of $\omegam$ and the
baryons in the \lya\ forest, and found that the values can be compatible
only for the extreme values of the parameters.  However, measurements of
$\Omega_{{\rm Ly} \alpha}$ are at present uncertain, so that it is
perhaps premature to imply that Machos are at odds with the amount of
baryons in the Ly$\alpha$ forest.  In addition, we have stressed the
difficulty in reconciling the Macho mass budget with the accompanying
carbon and/or nitrogen production in the case of white dwarfs.  The
overproduction of carbon or nitrogen by the white dwarf progenitors can
be diluted in principle, but this dilution would require even more
baryons that have not gone into stars.  At least in the simplest
scenario, in order not to conflict with the upper bounds on $\omegab$,
this would require an $\omegam$ slightly smaller than our lower limits
from extrapolating the Macho results.  Only 10$^{-2}$ of all baryons can
have passed through the white dwarf progenitors, a fraction that is in
conflict with our results for $\Omega_{{\rm prog}}$.

\section{Zero Macho Halo?}

The possibility exists that the 14 microlensing events that have been
interpreted as being in the Halo of the Galaxy are in fact due to some
other lensing population.  One of the most difficult aspects of
microlensing is the degeneracy of the interpretation of the data, so
that it is currently impossible to determine whether the lenses lie in
the Galactic Halo, or in the Disk of the Milky Way, or in the LMC.
Evans \etal (\cite{evans:diskflare}) proposed that the events could be
due to lenses in our own Milky Way Disk.  Gould (\cite{gouldy}) showed
that the standard model of the LMC does not allow for significant
microlensing.

Zhao (1998) has proposed that debris lying in a tidal tail stripped from
the progenitor of the LMC or SMC by the Milky Way or by an SMC-LMC tidal
interaction may explain the observed microlensing rate towards the
LMC. Within this general framework, he suggests that the debris thrown
off by the tidal interaction could also lead to a high optical depth for
the LMC.  There have been several observational attempts to search for
this debris.  Zaritsky \& Lin (1997) report a possible detection of such
debris in observations of red clump stars, but the results of further
variable star searches by the {\sc macho} group (Alcock \etal 1997b),
and examination of the surface brightness contours of the LMC (Gould
1998) showed that there is no evidence for such a population.  A stellar
evolutionary explanation for the observations of Zaritsky \& Lin (1997)
was proposed by Beaulieu \& Sackett (1998).  However, possible evidence
for debris within a few kpc of the LMC along the line of sight is
reported by the {\sc eros} group (Graff \etal in preparation).  These
issues are currently unclear and are under investigation by many groups.

Note that a recent microlensing event towards the SMC, MACHO-98-SMC-1,
was due to a binary lens.  In this case it was possible to clearly
identify that the lens is in the SMC and not in our Halo (Albrow
\cite{planet}).  Parallax analysis of event MACHO-97-SMC-1 shows that
this event also is likely to be in the SMC (Palanque-Delabrouille \etal
1998).  Analysis of the binary lensing event MACHO-LMC-9 shows that this
event lies in the LMC.  So far, all the events which can be located lie
in the Magellanic Clouds.  However, the cause of the remaining events of
the LMC remains ambiguous and awaits further observations.

\section{Conclusions}

Microlensing experiments have ruled out a large class of possible
baryonic dark matter components.  Substellar objects in the mass range
$10^{-7}\msun$ all the way up to $10^{-2}\msun$ are ruled out by the
experiments.  In this talk I discussed the heavier possibilities in the
range $10^{-2}\msun$ to a few $\msun$.  I showed that brown dwarfs and
faint stars are ruled out as significant dark matter components; they
contribute no more than 1\% of the Halo mass density.  White dwarfs and
neutron stars are also extremely problematic.  The chemical abundance
constraints are formidable.  The D and He$^4$ production by the
progenitors of the white dwarfs can be in agreement with observation for
low $\Omega_{WD}$ and an IMF sharply peaked at low masses $\sim 2\msun$.
Unless carbon is never dredged up from the stellar core (as has been
suggested by Chabrier \cite{chabriernew}), overproduction of carbon
and/or nitrogen is problematic.  The relative amounts of these elements
that is produced depends on Hot Bottom Burning, but both elements are
produced at the level of at least solar enrichment, in conflict with
what is seen in our Halo and in Ly$\alpha$ systems.  One must either
abandon stellar remnants as dark matter or argue that the debris have
remained hot and segregated from cooler neutral matter.  However, the
observations of TeV $\gamma$-rays from Mkn501 at z=0.034 restrict the
infrared background of the universe and hence the white dwarf
progenitors that would have produced infrared light.  In sum, we have a
constraint on the remnant density, $\Omega_{\rm WD} \leq (1-3) \times
10^{-3} h^{-1}$.

Hence, in conclusion, 

\begin{description}

\item{1.} Nonbaryonic dark matter in our Galaxy seems to be required, and

\item{2.} The nature of the Machos seen in microlensing experiments and
interpreted as the dark matter in the Halo of our Galaxy remains a
mystery.  Are we driven to primordial black holes (Carr 1994; Jedamzik
\cite{jedam}), nonbaryonic Machos (Machismos?), mirror Machos (Mohapatra
and Teplitz \cite{mohap}) or perhaps a no-Macho Halo?
\end{description}

\section{Acknowledgments}

We are grateful for the hospitality of the Aspen Center for Physics,
where part of this work was done.  DG acknowledges the financial support
of the French Ministry of Foreign Affairs' Bourse Chateaubriand.  KF
acknowledges support from the DOE at the University of Michigan.  The
work of BDF was supported in part by DOE grant DE-FG02-94ER-40823.


\section{References}

\begin{thebibliography}{99}

\bibitem{af} Adams, F.C., \& Fatuzzo, M. 1996, ApJ, 464, 256

\bibitem{planet} Albrow, M. D. \etal 1999, ApJ in press,
astro-ph/9807086

\bibitem{macho:1yr} Alcock, C., et al.\ 1996, ApJ, 461, 84

\bibitem{macho:2yr} \bline.\ 1997a, ApJ, 486, 697

\bibitem{macho:nodwf} \bline.\ 1997b, ApJ 490, L59

\bibitem{macho:smc} \bline.\ 1997c, ApJ, 491, L11

\bibitem{models96} Alexander, D. R., Brocato,
E., Cassisi, S., Castelliani, V., Ciaco, F., \& Degl'Innocenti,
S. 1996 submitted to \aap.

\bibitem{ansari} Ansari, R., et al. \ 1996, A\&A, 314, 94

\bibitem{axelrod} Axelrod, T., MACHO experiment,
talk presented at Aspen microlensing workshop, June 1997

\bibitem{bs} Bahcall, J.N. \& Soniera, R. 1980, ApJS, 44, 96

\bibitem{bfgk} Bahcall, J.N., Flynn, C., \& Gould, A.,
\& Kirhakos, S., 1994, ApJ, 435, L51

\bibitem{bc} Baraffe, I., Chabrier, G., Allard, F.,
and Hauschildt, P.H. 1995, \apj, {\bf 466}, L35.

\bibitem{bsac} Beaulieu J.-P. \& Sackett, P.D. 1997, AJ, 116, 209

\bibitem{btb94} Boeshaar, P. C.,
Tyson, J. A. \& Bernstein, G. M., 1994, \baas, {\bf 185}, \#22.02.

\bibitem{pilar} Canal, R., Isern, J., \% Ruiz-Lapuente, P.
1997, ApJ, 488, L35

\bibitem{carr} Carr, B. 1994, ARAA, 32, 531

\bibitem{chabriernew} Chabrier, G. 1999, ApJ Lett, in press;
astro-ph/9901145

\bibitem{cs} Charlot, S., \& Silk, J. 1995, ApJ, 445, 124

\bibitem{dahn} Dahn, C.C., Liebert, J., Harris, H.C.,
\& Guetter, H. H. 1995, in The Bottom of the Main Sequence and
Beyond, ed. C.G. Tinney (Heidelberg: Springer),
239.

\bibitem{dal} Dalcanton, J., Canizares, C., Granados, A.,
Steidel, C., \% Stocke, J., 1994, ApJ, 424, 550

\bibitem{evans} Evans, N.W. 1996, MNRAS, 278, L5

\bibitem{evans:diskflare} Evans, N.W., Gyuk, G., Turner, M.S.,
Binney, J. 1998, ApJ, 501, 45

\bibitem{ffg} Fields, B., Freese, K., \& Graff, D.,
1998, New Astron, 3, 347

\bibitem{ffg2} Fields, B., Freese, K., \% Graff, D., 1999,
submitted for publication, ApJ

\bibitem{fo98} Fields, B.D., \& Olive, K.A. 1998, ApJ, 506, 177

\bibitem{fms} Fields, B.D., Mathews, G.J., \& Schramm, D.N.,
1997, ApJ, 483, 625

\bibitem{fgb} Flynn, C., Bahcall, J., and Gould, A. 1996,
ApJ, 466, L55

\bibitem{freese} Freese,K., Fields, B., and Graff, D.,
``What are Machos? Limits on Stellar Objects as the Dark Matter
of Our Halo," Conf. Proc. of the International Workshop
on Aspects of Dark Matter in Astro and Particle Physics, Heidelberg,
Germany, astro-ph/9901178

\bibitem{fhp} Fukugita, M., Hogan, C.J., \& Peebles, P.J.E.
1997, ApJ, submitted (astro-ph/9712020)

\bibitem{ggt} Gates, E., Gyuk, G., \& Turner, M.S. 1996, PRD,
53, 4138

\bibitem{gght} Gates, E.I., Gyuk, G., Holder, G.P.,
\& Turner, M.S. 1997, ApJL, submitted (astro-ph/9711110)

\bibitem{GEG} Gyuk, G., Evans, W., \& Gates, E., 1998, ApJ,
502, L29

\bibitem{gm} Gibson, B.K., \& Mould, J.R., 1997, ApJ, 482, 98

\bibitem{gould} Gould, A. 1998, ApJ, 499, 728

\bibitem{gf96a} Graff, D.S., \& Freese, K. 1996a, ApJ, 456, L49

\bibitem{gf96b} \bline.\ 1996b, ApJ, 467, L65

\bibitem{gfwp} Graff, D.S., Freese, K., Walker, T.P., \& Pinsonneault,
M.H. 1999, accepted for publication, ApJ Lett, astro-ph/9903181

\bibitem{gg} Graff, D.S. \& Gardiner, in preparation

\bibitem{glf} Graff, D.S., Laughlin, G., \& Freese, K., 1998,
ApJ, 499, 7

\bibitem{gouldy} Gould, A., 1995, ApJ, 441, 77

\bibitem{ho} Hegyi, D.J., \& Olive, K.A. 1986, ApJ, 300, 492

\bibitem{jedam} Jedamzik, K. 1997, Phys. Rev. D, 55, 5871

\bibitem{eammon1} Kerins, E.J. 1997, A\&A, 322, 709

\bibitem{kerev} Kerins, E. \& Evans, N. W., 1998, ApJ, 507, 221

\bibitem{kt} Kirkman, D. \& Tytler, D. 1997, ApJ, 489, L123

\bibitem{konop} Konopelko, A.K., Kirk, J.G., Stecker, F.W.,
\& Mastichiadis, A. 1999, astro-ph/9904057

\bibitem{la} Larson, R. B. 1992 \mnras, {\bf 256}, 641.

\bibitem{latt} Lattanzio, J.C., 1989, ApJ, 344, L25

\bibitem{lattbooth} Lattanzio, J.C. \& Boothroyd, A.I., 1997
astro-ph/9705186

\bibitem{ldm} Liebert, J., Dahn, C. C., \& Monet, D.G. 1988,
ApJ, 332, 891

\bibitem{loeb} Loeb, A. 1997 (astro-ph/9704290)

\bibitem{lsbr} Lu, L., Sargent, W.L.W., Barlow, T.A., \& Rauch, M.
1998, AJ, submitted (astro-ph/9802189)

\bibitem{mcs} M\'era, D., Chabrier, G., \& Schaeffer, R.
1996, Europhys.\ Lett., 33, 327

\bibitem{mcs2} M\'era, D., Chabrier, G., \& Schaeffer, R.
1998, A\&A, 330 937

\bibitem{mohap} Mohapatra, R.N., \& Teplitz, V.L. 1999,
``Mirror Matter Machos", astro-ph/9902085

\bibitem{eros:smc} Palanque-Delabrouille, et al.\ 1998, A\&A, in press
(astro-ph/9710194)

\bibitem{pp}Price, N.M. and
Podsiadlowski, P. 1995 \mnras, {\bf 273} 1041.

\bibitem{rv}Renzini, A., \& Voli, M., 1981, A\&A, 94, 175

\bibitem{shull} Shull, M., \etal, 1998, AJ, 116, 2094

\bibitem{song} Songaila, A., Cowie, L.L., Hogan, C., \& Rugers, M.
1994, Nature, 368, 599

\bibitem{sc} Songaila, A., \& Cowie, L.L. 1996, AJ, 112, 335

\bibitem{wmhk} Weinberg, D.H., Miralda-Escud\'e, J.,
Hernquist, L., \& Katz, N. 1997, ApJ, 564

\bibitem{vdhg} van den Hoek, L.B., \& Groenewegen,
M.A.T. 1997, A\&AS, 123, 305

\bibitem{vot} Venkatesan, A, Olinto, A.V., \& Truran, J.W.
1999, ApJ in press (astro-ph/9705091)

\bibitem{zl} Zaritsky, D., \& Lin, D.N.C. 1997, AJ, 114, 2545

\bibitem{zhao} Zhao, H.S. 1998 MNRAS, 294, 139

\bibitem{z} Zinnecker, H. 1984 \mnras, {\bf 210}
43.

\bibitem{zuc} Zucca, E., et al.\ 1997, A\&A, 326, 477

\end{thebibliography}
\end{document}